# Inequalities for Non-Equilibrium Fluctuations of Work

Alexander Davydov
AlgoTerra LLC, 249 Rollins Avenue, Suite 202, Rockville, MD 20852
E-mail: alex.davydov@algoterra.com

December 25, 2010

**Abstract.** Five previously unknown inequalities relating equilibrium free energy differences and non-equilibrium work fluctuations are derived, and lucid path to derivation of many similar inequalities is presented. These results are based upon combined exploitation of the Jarzynski equality and the generalization of the scheme for producing uncertainty-type inequalities due to H. Weyl. The inequalities may possibly lead to better understanding of behavior of the equilibrium free-energy estimators from non-equilibrium experimental data in many important applications concerning biological, chemical, and physical molecular processes.

## 1. Introduction

Recent theoretical developments in non-equilibrium thermodynamics, such as establishment of the *Jarzynski equality* (JE) [1] and the *Crooks fluctuation theorem* [2], provide a remarkable link between equilibrium free energy differences and non-equilibrium fluctuations of work performed on the system. These relationships hold independently from the specific protocol of system perturbation and remain valid even in the far-from-equilibrium limit. In this paper, I present five previously unknown inequalities relating *equilibrium* free energy differences and *non-equilibrium* work fluctuations. Many other inequalities can be generated using the same approach. Their derivation is based on joint use of the JE and generalized version of the method due to H. Weyl [3] which can produce an unlimited number of uncertainty-type inequalities for a wide class of functions of a random variable. In addition, many inequalities relating various statistical averages for (equilibrium or non-equilibrium) work fluctuations can be generated by this technique, and few such inequalities are explicitly presented herein.

The paper is organized as follows. Section 2 presents the method of finding inequalities between various statistical averages of functions of a random variable. This method is general and its applicability extends far beyond the narrative of this paper. Section 3 briefly recalls the JE and its context preparing the reader for Section 4 where the main results (five inequalities) are obtained. Several additional relations involving work fluctuations are given in Section 5. Section 6 employs one of the derived inequalities to calculate lower bounds for the bias, variance, and mean square error (MSE) of the Jarzynski free energy difference estimator. Finally, Section 7 summarizes the results.

## 2. Generalization of the Weyl's Method

Consider a scalar random variable $X$ distributed according to a one-parameter probability density function (pdf) $p(x|\beta)$, where $x \in \Re, \beta \in [0, +\infty[$ with the normalization condition:

$$\int_{-\infty}^{+\infty} p(x|\beta)dx = 1 \qquad \forall \beta \geq 0 \qquad (1)$$

Let us introduce a real-valued 'amplitude' function, $\psi(x|\beta)$,

$$p(x|\beta) = \psi^2(x|\beta), \qquad (2)$$

and write down two obvious inequalities:

$$A \equiv \int_{-\infty}^{+\infty} (\alpha \cdot Q(x,\beta) \cdot \psi + \tfrac{\partial \psi}{\partial x})^2 \, dx \geq 0 \qquad (3)$$

$$B \equiv \int_{-\infty}^{+\infty} (\alpha \cdot Q(x,\beta) \cdot \psi + \tfrac{\partial \psi}{\partial \beta})^2 \, dx \geq 0 \qquad (4)$$

Here $\alpha$ denotes a real number and $Q(x,\beta)$ is some continuous function of its arguments with properties to be specified later. Note that both inequalities are satisfied for an arbitrary $\alpha \in \Re$. Let us start with the inequality (3), and, following the idea of Hermann Weyl (who considered only the particular case $Q(x,\beta) \equiv x$, see [3]), expand out the integral algebraically into a sum of three integrals as follows

$$A \equiv \alpha^2 \cdot I_1 + 2\alpha \cdot I_2 + I_3 \qquad (5)$$

$$I_1 \equiv \int_{-\infty}^{+\infty} Q^2 \psi^2 \, dx \qquad (5.\text{a})$$

$$I_2 \equiv \int_{-\infty}^{+\infty} Q\psi \cdot \frac{\partial \psi}{\partial x} \, dx = \tfrac{1}{2} \int_{-\infty}^{+\infty} Q \, d(\psi^2) = \tfrac{1}{2} Q \psi^2 \big|_{-\infty}^{+\infty} - \tfrac{1}{2} \int_{-\infty}^{+\infty} \psi^2 \, \tfrac{\partial Q}{\partial x} \, dx \qquad (5.\text{b})$$

$$I_3 \equiv \int_{-\infty}^{+\infty} \left(\tfrac{\partial \psi}{\partial x}\right)^2 dx = \tfrac{1}{4} \int_{-\infty}^{+\infty} p^{-1} \left(\tfrac{\partial p}{\partial x}\right)^2 dx \qquad (5.\text{c})$$

If we choose function $Q(x,\beta)$ such that (i) $Qp \to 0$ as $|x| \to \infty$, (ii) $\int_{-\infty}^{+\infty} pQ^2 \, dx < \infty$, and

(iii) $\int_{-\infty}^{+\infty} p \tfrac{\partial Q}{\partial x} \, dx < \infty$, then we obtain $I_1 = \langle Q^2 \rangle$ and $I_2 = -\tfrac{1}{2} \left\langle \tfrac{\partial Q}{\partial x} \right\rangle$. Integral $I_3$, up to a factor $\tfrac{1}{4}$, coincides with the *Fisher information* [4, 5] for the variable $x$, which we denote $I_x$. Thus, expression (5) can be recast in the form

$$A \equiv \alpha^2 \cdot \langle Q^2 \rangle - \alpha \cdot \left\langle \tfrac{\partial Q}{\partial x} \right\rangle + \tfrac{1}{4} I_x \qquad (6)$$

The fact that, for any real $\alpha$, $A \geq 0$ means that the discriminant of the corresponding quadratic equation $A(\alpha) = 0$ is non-positive, which leads to an inequality:

$$\langle Q^2 \rangle \cdot I_x \geq \left\langle \frac{\partial Q}{\partial x} \right\rangle^2, \qquad (7)$$

where

$$I_x \equiv \int_{-\infty}^{+\infty} p^{-1} \left(\frac{\partial p}{\partial x}\right)^2 dx = \left\langle \left(\frac{\partial \ln p}{\partial x}\right)^2 \right\rangle = 4 \int_{-\infty}^{+\infty} \left(\frac{\partial \psi}{\partial x}\right)^2 dx \qquad (8)$$

In the case $Q(x, \beta) \equiv x - \langle x \rangle$, (7) recovers the well-known *Cramer-Rao inequality* [4]

$$\mathrm{var}(x) \cdot I_x \geq 1, \qquad (9)$$

where $\mathrm{var}(x) \equiv \langle (x - \langle x \rangle)^2 \rangle = \langle x^2 \rangle - \langle x \rangle^2$.

Let us now examine the inequality (4) along the similar lines:

$$B \equiv \alpha^2 \cdot J_1 + 2\alpha \cdot J_2 + \tfrac{1}{4} I_\beta \qquad (10)$$

$$J_1 \equiv \int_{-\infty}^{+\infty} Q^2 \psi^2 dx = \langle Q^2 \rangle \qquad (10.\mathrm{a})$$

$$J_2 \equiv \int_{-\infty}^{+\infty} Q\psi \cdot \frac{\partial \psi}{\partial \beta} dx = \tfrac{1}{2}\left(\frac{\partial}{\partial \beta} \int_{-\infty}^{+\infty} Q\psi^2 dx - \int_{-\infty}^{+\infty} \frac{\partial Q}{\partial \beta} \psi^2 dx\right) = \tfrac{1}{2}\left(\frac{\partial}{\partial \beta}\langle Q \rangle - \left\langle \frac{\partial Q}{\partial \beta}\right\rangle\right) \qquad (10.\mathrm{b})$$

$$I_\beta \equiv 4\int_{-\infty}^{+\infty}\left(\frac{\partial \psi}{\partial \beta}\right)^2 dx = \int_{-\infty}^{+\infty} p^{-1}\left(\frac{\partial p}{\partial \beta}\right)^2 dx = \left\langle \left(\frac{\partial \ln p}{\partial \beta}\right)^2 \right\rangle \qquad (10.\mathrm{c})$$

As in the case before, we must impose certain constraints on function $Q(x, \beta)$ to ensure the existence of the integrals involved in (10). These constraints read: (i) $\int_{-\infty}^{+\infty} pQ^2 dx < \infty$, and (ii) $\int_{-\infty}^{+\infty} p \frac{\partial Q}{\partial \beta} dx < \infty$. The condition that $B(\alpha) \geq 0$ for any real $\alpha$ leads to an inequality

$$\langle Q^2 \rangle \cdot I_\beta \geq \left(\frac{\partial}{\partial \beta}\langle Q \rangle - \left\langle \frac{\partial Q}{\partial \beta}\right\rangle\right)^2 \qquad (11)$$

with $I_\beta$ being recognized as Fisher information for parameter $\beta$. This inequality can be recast in yet another form after we notice that $\frac{\partial}{\partial \beta}\langle Q \rangle - \langle \frac{\partial Q}{\partial \beta}\rangle = \int_{-\infty}^{+\infty} Q \frac{\partial p}{\partial \beta} dx = \langle Q \frac{\partial \ln p}{\partial \beta} \rangle$ and thus obtain

$$\langle Q^2 \rangle \cdot I_\beta \geq \left\langle Q \frac{\partial \ln p}{\partial \beta} \right\rangle^2 \tag{11.a}$$

It is worth noting that an alternative way of deriving inequalities (7), (11), and (11.a) relies on using the *Cauchy-Schwarz inequality* [6]

$$(f \cdot f) \cdot (g \cdot g) \geq [(f \cdot g)]^2 \tag{12}$$

with the choice $f = \sqrt{p} \cdot Q$; $g = \sqrt{p} \cdot \frac{\partial \ln p}{\partial x}$ to obtain (7) and $f = \sqrt{p} \cdot Q$; $g = \sqrt{p} \cdot \frac{\partial \ln p}{\partial \beta}$ to get (11.a) and then (11) as a consequence. However, the method used above, though less elegant, allowed us to keep the accurate track of all constraints on function $Q(x, \beta)$ which is crucial for further analysis.

## 3. Jarzynski Equality

Consider a classical system at equilibrium and in contact with a heat reservoir at temperature $T$. Let us force the system out of stationarity by changing its external parameter $\lambda(t)$ from $\lambda(t_1) = \lambda_A$ to $\lambda(t_1 + \tau) = \lambda_B$ during a finite interval $\tau$ and then let the system reach thermal equilibrium with the reservoir again, while holding the control parameter fixed at $\lambda = \lambda_B$. According to the second law of thermodynamics, the work performed on the system $W$ and the difference in free energy $\Delta F$ of the system between final and initial *equilibrium* configurations are related via the inequality

$$\langle W \rangle \geq \Delta F \tag{13}$$

with equality happening only in the case of a quasistatic process, i.e., when $\tau \to \infty$. The brackets in expression (13) denote an average over many realizations of the system perturbation, each time following exactly the same protocol when changing control parameter from $\lambda_A$ to $\lambda_B$. The work $W$ varies from one realization to the next due to unavoidable randomness in microscopic degrees of freedom of the system. The work pdf $P(W)$ depends on how the system is driven and is a functional of $\lambda(t)$. In 1997, C. Jarzynski has shown [1] that the distribution $P(W)$, regardless of the choice of external protocol $\lambda(t)$, obeys the equality (JE):

$$\langle e^{-\beta W} \rangle \equiv \int_{-\infty}^{+\infty} P(W) \cdot e^{-\beta W} dW = e^{-\beta \Delta F} \quad \forall \lambda(t) \tag{14}$$

Here $\beta \equiv 1/(k_B T)$ and $k_B$ is the Boltzmann's constant. Since its discovery, the JE has attracted significant attention and has been confirmed in both simulations and experiments. This remarkable relation between equilibrium free energy differences and non-equilibrium work fluctuations has become central in various applications [7].

## 4. Inequalities for Non-equilibrium Work Fluctuations

In the context of JE, let us choose the particular external protocol $\lambda(t)$ and define the corresponding distribution of work fluctuations as $P(W | \beta)$, where we explicitly point to possible dependence of the pdf upon *initial temperature T* of the system in thermal equilibrium with a heat reservoir ($\beta^{-1} = k_B T$). If we start the process of disturbing the system (using the same protocol $\lambda(t)$) at a slightly different temperature, corresponding to $\beta' = \beta + \delta\beta$ with small $\delta\beta$, then we expect the new distribution $P(W | \beta')$ to differ only a little from the old one. Mathematically, this means that we assume $P(W | \beta)$ to be (at the very least) a smooth function of two variables $W$ and $\beta$ of the class $\mathbf{C^1}$. The domain for $P(W | \beta)$ is set to be ($W \in \Re, \beta \in [\beta_1, \beta_2]$), where the interval of possible values of $\beta$ can be narrow. It is within this interval of temperatures that some of the inequalities derived below are valid. Physically, this implies that there are no phase transitions in the system within the interval $[\beta_1, \beta_2]$ which can potentially lead to drastic transformation of the shape of the distribution $P(W | \beta)$ after even small change of the initial temperature.

Looking at the inequalities (7), (11), and (11.a), it is easy to notice that, by proper choice of a function $Q(W, \beta)$ such that facilitates the JE (14), new relations between *equilibrium* free energy differences and *non-equilibrium* work averages can be established. In fact, there exist an unlimited number of possibilities for generating relations (inequalities) of this kind. Four most obvious choices of $Q(W, \beta)$ are listed below:

1. $Q(W, \beta) \equiv e^{-\beta W}$
2. $Q(W, \beta)^2 \equiv e^{-\beta W}, \Rightarrow Q \equiv e^{-\frac{1}{2}\beta W}$
3. $Q(W, \beta) \equiv W e^{-\beta W}, \Rightarrow \frac{\partial Q}{\partial W} \equiv e^{-\beta W} - \beta W e^{-\beta W}$
4. $Q(W, \beta) \cdot \frac{\partial \ln P}{\partial \beta} \equiv e^{-\beta W}, \Rightarrow Q \equiv e^{-\beta W} \left(\frac{\partial \ln P}{\partial \beta}\right)^{-1}$

Before we examine these possibilities one by one, notice that, if we make an assumption that distribution $P(W | \beta)$ decays faster than $\exp(-2\beta |W|)$ in the limit $W \to -\infty$, then all the constraints on properties of $Q(W, \beta)$ derived in Section 2 would be satisfied. So, in what follows, we postulate this feature of work distribution $P(W | \beta)$ without further deliberations.

Case #1:
$$Q(W, \beta) \equiv e^{-\beta W}, \Rightarrow \frac{\partial Q}{\partial W} = -\beta e^{-\beta W}, \quad \frac{\partial Q}{\partial \beta} = -W e^{-\beta W}$$

From inequality (11) combined with the JE (14) and the thermodynamic relation for the *equilibrium* mean energy $E = \frac{\partial}{\partial \beta}(\beta F)$, one obtains

$$|\Delta E \cdot e^{-\beta \Delta F} - \langle We^{-\beta W}\rangle| \leq \sqrt{I_\beta \cdot \langle e^{-2\beta W}\rangle} \tag{A}$$

with

$$I_\beta \equiv \int_{-\infty}^{+\infty} P^{-1}(W\mid\beta) \cdot \left(\frac{\partial P(W\mid\beta)}{\partial \beta}\right)^2 dW \quad [\text{Joules}^2] \tag{15}$$

From inequality (7) in combination with the JE, we arrive at

$$e^{-\beta \Delta F} \leq \beta^{-1} \cdot \sqrt{I_w \cdot \langle e^{-2\beta W}\rangle} \tag{B2}$$

with

$$I_w \equiv \int_{-\infty}^{+\infty} P^{-1}(W\mid\beta) \cdot \left(\frac{\partial P(W\mid\beta)}{\partial W}\right)^2 dW \quad [\text{Joules}^{-2}] \tag{16}$$

Case #2:
$$Q(W,\beta) \equiv e^{-\frac{1}{2}\beta W}, \Rightarrow \frac{\partial Q}{\partial W} = -\frac{1}{2}\beta e^{-\frac{1}{2}\beta W}, \quad \frac{\partial Q}{\partial \beta} = -\frac{1}{2}We^{-\frac{1}{2}\beta W}$$

Inequality (7) together with the JE leads to

$$e^{-\beta \Delta F} \geq \frac{\beta^2}{4I_w}\langle e^{-\frac{1}{2}\beta W}\rangle^2 \tag{B1}$$

From (11) and the JE, we obtain

$$e^{-\beta \Delta F} \geq I_\beta^{-1} \cdot \left[\tfrac{1}{2}\langle We^{-\frac{1}{2}\beta W}\rangle + \tfrac{\partial}{\partial \beta}\langle e^{-\frac{1}{2}\beta W}\rangle\right]^2 \tag{C}$$

From (11.a) and the JE, we get

$$e^{-\beta \Delta F} \geq I_\beta^{-1} \cdot \langle e^{-\frac{1}{2}\beta W} \cdot \tfrac{\partial}{\partial \beta}(\ln P)\rangle^2 \tag{D1}$$

Case #3:
$$Q(W,\beta) \equiv We^{-\beta W}, \Rightarrow \frac{\partial Q}{\partial W} = e^{-\beta W} - \beta We^{-\beta W}, \quad \frac{\partial Q}{\partial \beta} = -W^2 e^{-\beta W}$$

From inequality (7) and the JE, one obtains

$$|e^{-\beta\Delta F} - \beta\langle We^{-\beta W}\rangle| \leq \sqrt{I_w \cdot \langle W^2 e^{-2\beta W}\rangle} \qquad (E)$$

Case #4:
$$Q(W,\beta) \equiv e^{-\beta W}\left(\tfrac{\partial \ln P}{\partial \beta}\right)^{-1}$$

From (11.a) and the JE, we find

$$e^{-\beta\Delta F} \leq \sqrt{I_\beta \cdot \left\langle \frac{e^{-2\beta W}}{\left(\frac{\partial \ln P}{\partial \beta}\right)^2}\right\rangle} \qquad (D2)$$

It is convenient to arrange all results obtained in this Section into Table 1 where I have combined single inequalities into double ones whenever possible.

**Table 1**: Summary of results

| Reference label | *Inequalities for non-equilibrium work fluctuations* |
|---|---|
| A | $\|\Delta E \cdot e^{-\beta\Delta F} - \langle We^{-\beta W}\rangle\| \leq \sqrt{I_\beta \cdot \langle e^{-2\beta W}\rangle}$ |
| B | $\dfrac{\beta^2}{4I_w} \cdot \langle e^{-\tfrac{1}{2}\beta W}\rangle^2 \leq e^{-\beta\Delta F} \leq \beta^{-1} \cdot \sqrt{I_w \cdot \langle e^{-2\beta W}\rangle}$ |
| C | $I_\beta^{-1} \cdot \left[\tfrac{1}{2}\langle We^{-\tfrac{1}{2}\beta W}\rangle + \tfrac{\partial}{\partial\beta}\langle e^{-\tfrac{1}{2}\beta W}\rangle\right]^2 \leq e^{-\beta\Delta F}$ |
| D | $I_\beta^{-1} \cdot \langle e^{-\tfrac{1}{2}\beta W} \tfrac{\partial}{\partial\beta}(\ln P)\rangle^2 \leq e^{-\beta\Delta F} \leq \sqrt{I_\beta \cdot \left\langle \dfrac{e^{-2\beta W}}{\left(\tfrac{\partial \ln P}{\partial \beta}\right)^2}\right\rangle}$ |
| E | $\|e^{-\beta\Delta F} - \beta\langle We^{-\beta W}\rangle\| \leq \sqrt{I_w \cdot \langle W^2 e^{-2\beta W}\rangle}$ |

## 5. Some Supplementary Inequalities

Many additional inequalities can be obtained using the general formulae (7) and (11) derived in Section 2. These relations may or may not allow for the substitution of the JE to derive lower and upper bounds for $e^{-\beta\Delta F}$ as shown earlier. However, they can be of interest in their own right. For instance, if we select $Q(W,\beta) \equiv W^n$, with $n = 1, 2, ...$, then, from (7) and (11), we obtain two potentially useful inequalities, respectively:

$$\langle W^{2n}\rangle \geq I_w^{-1} n^2 \langle W^{n-1}\rangle^2 \tag{17}$$

and

$$\langle W^{2n}\rangle \cdot I_\beta \geq \left[\tfrac{\partial}{\partial \beta}\langle W^n\rangle\right]^2, \tag{18}$$

which can be combined into a single inequality

$$\langle W^{2n}\rangle \geq \max\left\{I_w^{-1}\cdot n^2\langle W^{n-1}\rangle^2,\ I_\beta^{-1}\cdot\left(\tfrac{\partial}{\partial \beta}\langle W^n\rangle\right)^2\right\} \tag{19}$$

Similarly, by choosing $Q(W,\beta) \equiv (W-\langle W\rangle)^n$, with $n=1,2,\ldots$, we get

$$\langle (W-\langle W\rangle)^{2n}\rangle \geq \max\left\{I_w^{-1}\cdot n^2\langle (W-\langle W\rangle)^{n-1}\rangle^2,\ I_\beta^{-1}\cdot\langle (W-\langle W\rangle)^n \tfrac{\partial}{\partial \beta}(\ln P)\rangle^2\right\} \tag{20}$$

or, alternatively,

$$\langle (W-\langle W\rangle)^{2n}\rangle \geq \max\left\{I_w^{-1}\cdot n^2\langle (W-\langle W\rangle)^{n-1}\rangle^2,\ I_\beta^{-1}\cdot\left(\tfrac{\partial}{\partial \beta}\langle (W-\langle W\rangle)^n\rangle + n\tfrac{\partial \langle W\rangle}{\partial \beta}\langle (W-\langle W\rangle)^{n-1}\rangle\right)^2\right\} \tag{21}$$

Putting $n=1$ in (21), we obtain the lower bound for the variance of $W$:

$$\mathrm{var}(W) \geq \max\left\{I_w^{-1},\ I_\beta^{-1}\cdot\left(\tfrac{\partial}{\partial \beta}\langle W\rangle\right)^2\right\} \tag{22}$$

## 6. Possible Applications

In this Section, I present an example showing how one of the inequalities derived in Section 4 can be exploited to improve the Jarzynski free energy difference estimator by finding the lower bounds for its bias, variance, and MSE. Finding free energy differences have a variety of important applications in physical, chemical, and biological systems. Examples include drug design, estimates of chemical potentials and of binding affinities of ligands to proteins, determination of the solubility of small molecules, etc (e.g., see [8]). The JE provides a convenient means to estimate the free energy change between an initial equilibrium state and a second equilibrium state which is reached by a non-equilibrium process. If $N$ perturbations of a system are performed using the same protocol, the Jarzynski free energy difference estimator is given by [1]

$$\Delta \hat{F}_J(N) = -\beta^{-1}\ln\left[\tfrac{1}{N}\sum_{i=1}^{N} e^{-\beta W_i}\right], \tag{23}$$

where $W_i$ denotes the work spent in changing the state of the system during the $i$-th experiment. Unfortunately, as a result of exponential averaging that strongly depends on the behavior at the tails of the distribution, the Jarzynski estimator is inherently noisy and biased for all finite $N$. However, the bias decreases monotonically with increasing $N$ and vanishes in the limit $N \to \infty$ [9]. The detailed analysis of the statistical properties on the Jarzynski estimator has been given in [10] where the estimates for its bias, variance, and MSE were obtained in various regimes. In particular, it was shown that, in the large $N$ limit and for arbitrary perturbations, the following relations hold true

$$\text{Bias} \qquad B_J(N) \equiv \langle \Delta \hat{F}_J(N) \rangle - \Delta F \cong \frac{\text{var}(e^{-\beta W})}{2\beta N e^{-2\beta \Delta F}} \qquad (24)$$

$$\text{Variance} \qquad \sigma_J^2(N) \equiv \langle (\Delta \hat{F}_J(N) - \langle \Delta \hat{F}_J(N) \rangle)^2 \rangle \cong \frac{2B_J(N)}{\beta} \qquad (25)$$

$$\text{Mean Square Error} \quad MSE_J(N) \equiv \langle (\Delta \hat{F}_J(N) - \Delta F)^2 \rangle = \sigma_J^2(N) + B_J^2(N) \approx \sigma_J^2(N) \quad (26)$$

Note that

$$\text{var}(e^{-\beta W}) = \langle e^{-2\beta W} \rangle - \langle e^{-\beta W} \rangle^2 = \langle e^{-2\beta W} \rangle - e^{-2\beta \Delta F}, \qquad (27)$$

where the JE was employed in the last transformation. From the inequality (B), it follows that

$$\langle e^{-2\beta W} \rangle \geq \frac{\beta^2}{I_w} e^{-2\beta \Delta F} \qquad (28)$$

Combining (24), (27), and (28), one obtains the lower bounds for bias, variance, and MSE of the Jarzynski estimator

$$B_J(N) \geq \frac{1}{2\beta N} \left( \frac{\beta^2}{I_w} - 1 \right) \qquad (29)$$

$$\sigma_J^2(N) \geq \frac{1}{N} \left( \frac{1}{I_w} - \frac{1}{\beta^2} \right) \qquad (30)$$

$$MSE_J(N) \geq \frac{1}{N} \left( \frac{1}{I_w} - \frac{1}{\beta^2} \right) + \frac{1}{4\beta^2 N^2} \left( \frac{\beta^2}{I_w} - 1 \right)^2 \qquad (31)$$

The above bounds are sensible only if $I_w \leq \beta^2$. Since the Cramer-Rao inequality [4] yields $I_w \geq 1/\sigma_w^2$, we observe that necessary (but not sufficient) condition for the relations (29)-(31) to be useful reads $\sqrt{\sigma_w^2} \geq k_B T$, which is likely the case in most experimental situations. Knowing the lower bound for the bias (29) allows one to improve the Jarzynski estimator by subtracting this quantity from the estimated free energy difference. The result is the *partially bias-corrected* (*pbc*) Jarzynski estimator:

$$\Delta \hat{F}_J^{pbc}(N) = \begin{cases} \Delta \hat{F}_J(N), & I_w > \beta^2 \\ \Delta \hat{F}_J(N) - \frac{1}{2\beta N}(\frac{\beta^2}{I_w} - 1), & I_w \leq \beta^2 \end{cases} \quad (32)$$

If the analytic form of pdf $P(W)$ for work spent on changing the state of the system is unknown, finding the corresponding Fisher information $I_w$ from empirical data is a challenging task. One possible approach to estimating $I_w$ is based on utilization of the amplitude function introduced earlier in Section 2. Indeed, one can recast the expression (16) as

$$I_w = 4 \int_{-\infty}^{+\infty} \left( \frac{\partial \sqrt{P(W)}}{\partial W} \right)^2 dW \approx \frac{4}{(\Delta W)^2} \sum_{i=1}^{M-1} \left( \sqrt{h_{i+1}} - \sqrt{h_i} \right)^2 \quad (33)$$

where $h_k$ denotes the value of the normalized histogram for work $W$ at bin $k$, $M$ is the total number of bins of the same size $\Delta W$, and $1 << M << N$. How well this approximation works in practice remains to be investigated. It is worth mentioning, however, that, when distribution $P(W)$ is Gaussian, the estimation of $I_w$ becomes trivial. The Cramer-Rao inequality transforms into *equality* in this case and thus yields: $I_w = 1/\sigma_w^2$.

## 7. Conclusions

The method for generating an unlimited number of inequalities relating equilibrium free energy differences and non-equilibrium fluctuations of work performed on a system has been presented. As an illustration of the method's power, five previously unknown inequalities of that sort have been explicitly obtained. The method relies on combined use of the Jarzynski equality and generalized version of the method due to H. Weyl which can produce the uncertainty-type inequalities for a wide class of functions of a random variable. These inequalities offer the lower and/or upper bounds for the equilibrium free energy differences and potentially may result in better understanding of the behavior of equilibrium free-energy estimators from non-equilibrium experimental data in many important applications concerning biological, chemical, and physical molecular processes.

**Acknowledgements**

The author is grateful to Mikhail Anisimov and Chris Jarzynski for encouragement and interest in this work.